\documentclass[twocolumn,pra,showpacs,final,superscriptaddress,10pt,aps,floatfix]{revtex4}
\usepackage[usenames]{color}
\usepackage{amsmath}
\usepackage{amssymb}
\usepackage{mathrsfs}
\usepackage{bbold}
\usepackage{euscript}
\usepackage{epic, eepic}
\usepackage{subfigure}
\usepackage{rotating}
\usepackage{latexsym}
\usepackage{amssymb}
\usepackage{graphicx}
\usepackage{graphics}
\usepackage{bm}

\begin{document}




\title{An efficient basis set representation for calculating electrons in molecules}
\author{Jeremiah R. Jones}
\affiliation{Department of Mathematics, Arizona State University, Tempe, Arizona 85281}
\author{Fran\c{c}ois-Henry Rouet}
\affiliation{Computing Sciences, Lawrence Berkeley National Laboratory, Berkeley CA 94720}
\author{Keith V. Lawler}
\affiliation{Department of Chemistry, University of Nevada-Las Vegas, Las Vegas, Nevada 89154}
\author{Eugene Vecharynski}
\affiliation{Computing Sciences, Lawrence Berkeley National Laboratory, Berkeley CA 94720}
\author{Khaled Z. Ibrahim}
\affiliation{Computing Sciences, Lawrence Berkeley National Laboratory, Berkeley CA 94720}
\author{Samuel Williams}
\affiliation{Computing Sciences, Lawrence Berkeley National Laboratory, Berkeley CA 94720}
\author{Brant Abeln}
\affiliation{Department of Chemistry, University of California, Davis, Davis CA 95616}
\affiliation{Ultrafast X-Ray Science Laboratory, Lawrence Berkeley National Laboratory, Berkeley CA 94720}
\affiliation{Chemical Sciences, Lawrence Berkeley National Laboratory, Berkeley CA 94720}
\author{Chao Yang}
\affiliation{Computing Sciences, Lawrence Berkeley National Laboratory, Berkeley CA 94720}
\author{Daniel  J. Haxton}
\affiliation{Ultrafast X-Ray Science Laboratory, Lawrence Berkeley National Laboratory, Berkeley CA 94720}
\affiliation{Chemical Sciences, Lawrence Berkeley National Laboratory, Berkeley CA 94720}
\author{C. William McCurdy}
\affiliation{Department of Chemistry, University of California, Davis, Davis CA 95616}
\affiliation{Ultrafast X-Ray Science Laboratory, Lawrence Berkeley National Laboratory, Berkeley CA 94720}
\affiliation{Chemical Sciences, Lawrence Berkeley National Laboratory, Berkeley CA 94720}
\author{Xiaoye S. Li}
\affiliation{Computing Sciences, Lawrence Berkeley National Laboratory, Berkeley CA 94720}
\author{Thomas N. Rescigno}
\affiliation{Chemical Sciences, Lawrence Berkeley National Laboratory, Berkeley CA 94720}

\begin{abstract}

The method of
McCurdy, Baertschy, and Rescigno,
J. Phys. B, \textbf{37}, R137 (2004)
%
%
is generalized to obtain a 
 straightforward, surprisingly accurate, 
and scalable numerical representation for calculating the electronic wave functions of molecules.
It uses a basis set of product sinc functions arrayed on a Cartesian grid, and yields
1 kcal/mol precision for valence transition energies with a grid resolution of approximately 0.1 bohr.
The Coulomb matrix elements are replaced with matrix elements obtained from the kinetic energy operator.
A resolution-of-the-identity approximation renders the
primitive one- and two-electron matrix elements diagonal; 
in other words, the Coulomb operator is local with respect
to the grid indices.
The calculation of contracted two-electron matrix elements among orbitals requires only
O($N \log(N)$) multiplication operations, not O($N^4$),  
where $N$ is the number of basis functions; $N=n^3$ on cubic grids.
The representation not only is numerically expedient, but also produces
energies and properties
superior to those calculated variationally.  
Absolute energies, absorption cross sections, transition energies, and ionization potentials are reported for
one- (He$^+$, H$_2^+$), two- (H$_2$, He), ten- (CH$_4$) and 56-electron (C$_8$H$_8$) systems.

 \end{abstract}

\maketitle


\section{Introduction} \label{section_one}

The inherent problem in scaling electronic structure methods to larger systems is the prohibitive cost of 
storing and transforming two-electron matrix elements,
which we denote in chemists' notation
\begin{equation}
[ij | kl] = \int\int d^3\vec{r}_1 d^3\vec{r}_2 \ \chi_i(\vec{r}_1) \chi_j(\vec{r}_1) \frac{1}{|\vec{r}_1 - \vec{r}_2|} \chi_k(\vec{r}_2) \chi_l(\vec{r}_2)
\label{eqone}
\end{equation}
for a basis $\{ \chi_i \}$.  The set of two-electron matrix elements is a fourth-rank tensor, such that 
transformations of the set require O(N$^4$) multiplication operations;
sophisticated methods such as coupled cluster must cope with even poorer scaling, O(N$^6$).  
There has been much work to circumvent this basic
problem~\cite{tensor1,tensor2,tensor3,tensor4,tensor5}, especially by Martinez and coworkers. 


We describe a basis set method for electronic structure motivated by the discrete variable representation (DVR) 
\cite{dickcert,lill+parker+light,light+hamilton+lill,corey+lemoine,leforestier,
coreytromp,slight,szalay,smeyer,littlejohn+cargo,yu}
%
%
that is an adaptation of the method in
Ref. \cite{topicalreview} to Cartesian coordinates.  
Starting with an evenly spaced grid basis
%
%
means that the number of basis functions is large, but also that the rank of the two-electron matrix element tensor is
automatically reduced from four to three due to redundancy.
Going further, using the generalization of Ref.~\cite{topicalreview}, we obtain a diagonal two-electron matrix element tensor --
in other words, 
we further reduce the tensor to the minimum rank one,
\begin{equation}
[ij | kl] \sim \delta_{ij}\delta_{kl} \ v_{i-k} \quad .
\label{finalexpr0}
\end{equation}
(In this equation the indices $i$,$j$,$k$, and $l$ each would run from 1 to $N = n^3$ on a cubic grid.)

Using this resolution-of-the-identity approximation for the treatment of the Coulomb
potential within the discrete variable representation, and 
employing established Fourier methods for triple Toeplitz linear algebra~\cite{davis1979circulant,chen2013parallelizing},
the computation of two-electron matrix elements among molecular orbitals takes O($N \log(N)$) time, not O($N^4$).
The method is therefore not quite ``linear-scaling'', but it is numerically exact;
it does not involve any truncated sums in a multipole expansion, for instance.

Gaussian basis sets have traditionally been the preferred single-electron representation for real-space 
electronic structure calculations, due to the localized nature of these functions and the speed with 
which matrix elements among them may be evaluated.
Although Gaussians have been widely successful,
they have inherit limitations in their flexibility; in particular, they are unable to represent electrons in the 
continuum, which is necessary for ionization and electron scattering applications. Furthermore, it is not 
always clear exactly how to obtain rigorous error bounds of basis set truncation.

There has recently been an increased interest among researchers in the field to develop grid-based 
methods using strictly numerical techniques that can handle a wider variety of problems and can be 
subjected to systematic error analysis. A thorough review of grid methods in electronic structure can be found in~\cite{torsti+others}.  
Some examples of grid-based techniques currently in use are 
finite differences~\cite{seisonen+puska+nieminen,ancilotto+blandin+toigo,chelikowsky+troullier+saad}, 
finite elements~\cite{white+wilkins+teter,pask+klein+fong+sterne}, and wavelets~\cite{arias}.
These methods make the treatment of arbitrary boundary conditions considerably easier than basis set methods. 
Another advantage of grid methods is the flexibility allowed in performing calculations on complicated 
spatial domains. Finite difference methods are limited in this regard since they require strictly rectangular 
meshes, whereas finite element methods offer complete freedom in choosing a computational mesh.

Similar to finite element methods, discrete variable representation (DVR) methods 
have characteristics of both a basis set method and 
a grid method in the sense that each basis function is localized around a specific grid point, 
and potential functions are evaluated as local multiplicative operators on the grid.
Many DVR bases have appeared in the 
literature, including those based on
Bessel functions \cite{littlejohn+cargo}
Lagrange polynomials \cite{yip+mccurdy+rescigno,rescigno+mccurdy},
 and sinc functions \cite{colbert+miller,jordan+mazziotti,mazziotti}, as well as
 multidimensional bases~\cite{smeyer} and others described in 
 Refs. \cite{dickcert,lill+parker+light,light+hamilton+lill,corey+lemoine,leforestier,
coreytromp,slight,szalay,smeyer,littlejohn+cargo,yu}. 

One issue in evaluating potential energy matrix elements for molecular systems 
is how to resolve the singularities that occur in the Coulomb potential terms. 
A number of methods for doing so have appeared in the literature, 
including the use of energy cut-off functions \cite{fusti+pulay,fusti} and 
multipole expansions via Legendre polynomials in spherical 
coordinates \cite{topicalreview, scherbinin+pupyshev+stepanov}.  
The singular Coulomb potential cannot be used straightforwardly within the DVR approximation,
because doing so would entail the use of infinite diagonal matrix elements.

To address this issue, the present method makes use of the fact that the 
Green's function for the Laplace operator is the Coulomb potential. 
In doing so it follows the derivation used in Ref.~\cite{topicalreview}.  
That work presented a treatment in spherical
polar coordinates, using the partial-wave expansion of the Green's function,
arriving at expressions for the two 
electron matrix elements diagonal in the radial index, corresponding to an 
expansion in Gauss-Lobatto DVR for the radial degree of freedom.  Here we do not use the
partial-wave expansion and instead treat Poisson's equation in Cartesian coordinates.

We present the method in the next section, then results, and finally
in the conclusion
we speculate about possible elaborations to the method that could make it
more versatile for excited state and time-dependent problems, and perhaps even
competitive with Gaussian basis sets for the computation of 
results requiring chemical accuracy~\cite{chemical_accuracy}.


\section{Method} \label{section_two}

\subsection{Sinc Basis}

Sinc functions have been used extensively in several areas of applied 
 mathematics including numerical solutions to ordinary and partial differential equations, 
 interpolation and Fourier analysis \cite{stenger} but were first introduced 
 in the context of a DVR basis for solving the Schr\"odinger equation by Colbert and Miller in \cite{colbert+miller};
 a good description of the sinc DVR can also be found in Ref.~\cite{groenenboom+colbert}.
 Sinc basis functions have been used in electronic structure in Refs.~\cite{colbert+miller,jordan+mazziotti,mazziotti}.

The sinc function is defined as
\begin{equation} 
\label{eq:sinc}
\textnormal{sinc}(x) =  \bigg\{\begin{array}{cc}  \frac{\sin(\pi x)}{\pi x} 
& \textnormal{if  } x \neq 0 \\ 1 & \textnormal{if } x = 0 \end{array}.  
\end{equation}
and an orthonormal basis in one dimension is
\begin{equation}
\label{eq:basis1D}
\xi_i(x) = \frac{1}{\sqrt{\Delta}}\textnormal{sinc}\left(\frac{x-x_i}{\Delta}\right)
\end{equation}
with $\Delta$ the uniform grid spacing, $x_{i+1} = x_i + \Delta$.

\subsection{Kinetic energy matrix elements}

The kinetic energy matrix elements among these functions are 
\begin{eqnarray}
\label{eq:kinetic1D}
t_{ij} &=& \bigg\langle \xi_i \bigg| -\frac{1}{2}\frac{d^2}{dx^2} \bigg| \xi_j \bigg\rangle \nonumber \\ &=&
\bigg\{\begin{array}{ll} \pi^2/(6\Delta^2) & \textnormal{if  } i = j \\ (-1)^{i-j}/(\Delta^2(i-j)^2) & \textnormal{if } i \neq j \end{array}.
\end{eqnarray}
Notice that these matrix elements only depend on $i-j$, i.e., $t$ is constant along diagonals, i.e., $t$ is Toeplitz.
A derivation of these elements is given in Ref. \cite{colbert+miller}. 
We make a three dimensional product basis in the straightforward way,
\begin{equation}
\chi_{\vec{i}}(x,y,z) = \xi_{i1}(x) \xi_{i2}(y) \xi_{i3}(z)  \quad .
\label{straightforward}
\end{equation}

The three dimensional kinetic energy is, as usual, 
\begin{equation}
T_{\vec{i}\vec{j}} = t_{i1,j1} \delta_{i2,j2} \delta_{i3,j3}  + \delta_{i1,j1} t_{i2,j2} \delta_{i3,j3} + \delta_{i1,j1} \delta_{i2,j2} t_{i3,j3}
\label{asusual}
\end{equation}
Since $t$ and the identity matrix are Toeplitz, $T$ is triple Toeplitz, as are 
the matrix elements of any translationally invariant operator.  We only use explicit vector-index notation in
Eqs.~\ref{straightforward} and~\ref{asusual}, and in sections~\ref{kesect} and~\ref{toepsect}.  
In the rest of this paper, we use contracted indices, such that for a three dimensional
basis function $\chi_i(\vec{r})$, or matrix element $T_{ij}$, 
the index $i$ (or $j$) represents a single integer that runs from 1 to $N=n^3$ on a cubic grid.

\subsection{Discrete Variable Representation resolution of the identity for two-electron matrix elements}

In the generalization of Ref.~\cite{topicalreview} to Cartesian coordinates, there are several simplifications
that result from the use of sinc basis functions.  The present method and that of Ref.~\cite{topicalreview}
are founded on the replacement of Coulomb matrix elements by matrix elements obtained from the kinetic energy
operator via a resolution-of-the identity approximation invoking Poisson's equation.  
However, for the two-electron matrix elements, it is not necessary to introduce the kinetic energy
operator into the derivation, if the sinc DVR is used.  Therefore, in this section, 
we provide the simplest derivation of the two-electron matrix elements
used in this method, before introducing the kinetic energy operator in the sections below.

The method of Ref.~\cite{topicalreview} 
uses the fact that the Coulomb potential is the Green's function of the Laplace operator
to avoid the inherent problem with using the discrete variable representation (DVR)
approximation for singular potentials.
It results in an expression, 
Eq.(\ref{finalexprfull}), which for the sinc
DVR basis is equivalent to the resolution of the identity described in this section:
\begin{equation}
[ij | kl] = 2\pi \delta_{ij}\delta_{kl}  (w_i w_j)^{-\frac{1}{2}} T_{ik}^{-1}  \nonumber
\end{equation}
where the $w$ are the quadrature weights -- presently, for the 3D Cartesian sinc basis uniformly equal to $\Delta^3$ -- 
and $T^{-1}$ is the limit of the inverse of the kinetic
energy matrix as the size of the basis is taken to be infinity.
Because the sinc basis is complete
in momentum space up to a cutoff, the matrix element of the matrix inverse is equal to the matrix 
element of the operator inverse,
\begin{equation}
(T^{-1})_{ik} = \int d^3r_1 d^3r_2 \ \chi_i(\vec{r}_1) \frac{1}{2 \pi \vert\vec{r}_{12}\vert} 
\chi_k(\vec{r}_2) \ .
\label{tinva}
\end{equation}

Because Eq.~\ref{tinva} holds for the sinc DVR basis, there is no need to introduce the kinetic
energy matrix into the derivation of the two-electron matrix elements.  Our final expression for them, 
Eq.~\ref{finalexpr}, results simply from a resolution-of-the-identity approximation.
                                   
The resolution of the identity makes straightforward use of the interpolating 
property of discrete variable representation
(DVR) basis functions: each basis function
belongs to a grid index, and is zero at all the grid points other than that corresponding to its own index.  
This ``discrete
orthogonality'' condition~\cite{lebedev_dvr} is the defining property of a DVR basis set.  
An arbitrary function can be expanded easily in such an 
interpolating basis,
\begin{equation}
f(x) \approx (w_i)^{-1/2} \sum_i  \chi_i(x) f(x_i)
\label{interp00}
\end{equation}
where $w_i$ is the quadrature weight at point $i$; presently the weights are all equal to $\Delta$ for
the one-dimensional sinc DVR and $\Delta^3$ for the 3D product basis.
This may be written as a resolution of the identity,
\begin{equation}
f(x) \approx \sum_i \vert \chi_i \rangle\langle \chi_i \vert \ f(x)
\end{equation}
where the integral is performed using the underlying quadrature, giving
\begin{equation}
\chi_k(\vec{r}) \chi_l(\vec{r}) \approx \delta_{kl} (w_k)^{-1/2} \chi_k(\vec{r}) 
\label{interp}
\end{equation}
 such that the density (a sum of squares of localized basis functions) is re-expanded as a 
sum of localized basis functions, without the square.
 The fact that the auxiliary basis, the one in which the density is expanded,
is the same as the basis in which the wave function is resolved means that, at least aesthetically, 
it is the \textit{ideal} resolution of the identity.  

The expression for the two-electron integral is obtained simply from Eqs.~\ref{eqone} and~\ref{interp},
\begin{equation}
[\vec{i}\vec{j} || \vec{k} \vec{l}]  \approx \Delta^{-3}
\delta_{ij}\delta_{kl} 
\int d^3r_1 d^3r_2 \ 
\chi_i(\vec{r}_1) \frac{1}{\vert\vec{r}_{12}\vert} \chi_k(\vec{r}_2) 
%
%
\label{finalexpr}
\end{equation}

\subsection{Application of the method of Ref.~\cite{topicalreview} to arbitrary three dimensional discrete variable representations\label{tnrsect}}

Here we provide a complete derivation of both the one- and two-electron matrix elements that follows
the derivation in Ref.~\cite{topicalreview} closely.  This method is founded upon 
the observation that the Coulomb potential is the Green's function of the kinetic energy (Laplace)
operator, and replaces Coulomb matrix elements in a basis with matrix elements obtained from the
kinetic energy operator in the same basis.
There are only two significant differences between the derivation in Ref.~\cite{topicalreview} and this one: one,
we use the full three-dimensional Green's function for the Laplacian operator, not its partial
wave expansion; and two, we eliminate the need for an explicit boundary condition.  
For sinc basis functions, the derivation of the two-electron matrix elements may be simplified as in the section above,
but the derivation here is applicable to general discrete variable representations in three dimensions
and includes both the one- and two-electron matrix elements.

We define 
\begin{equation}
\label{eq:ykl}
y^{kl}(\vec{r}_1) = \int d^3\vec{r}_2 \ \chi_k(\vec{r}_2) \chi_l(\vec{r}_2) \frac{1}{|\vec{r}_1 - \vec{r}_2|}
\end{equation}
so that, with reference to Eq.(\ref{eqone}), we can write
\begin{equation}
\label{eq:twoelectronsimple}
[ij | kl] = \int d^3\vec{r}_1  \ \chi_i(\vec{r}_1) \chi_j(\vec{r}_1) y^{kl}(\vec{r}_1) \quad .
\end{equation}
Applying the Laplacian to both sides of Eq.(\ref{eq:ykl}) results in the Poisson equation
\begin{equation}
\label{eq:poisson}
\nabla_{\vec{r}_1}^2 y^{kl}(\vec{r}_1) = -4\pi \chi_{k}(\vec{r}_1) \chi_l(\vec{r_1}) \quad .
\end{equation}
Here we have used the fact that
\begin{equation}
G(\vec{r},\vec{r}') \equiv -\frac{1}{4\pi|\vec{r}-\vec{r}'|}
\end{equation}
is the free space Green's function for the Laplace operator satisfying
\begin{equation}
\nabla^2_{\vec{r}'}G(\vec{r},\vec{r}') = \delta(\vec{r}-\vec{r}')
\end{equation}
with the boundary condition $G(\vec{r},\vec{r}') \rightarrow 0$ as $|\vec{r}| \rightarrow \infty$. 
We now approximate $y^{kl}(\vec{r}_1)$ as a linear combination of the basis functions,
\begin{equation}
\label{yklexpand}
y^{kl}(\vec{r}_1) \approx \sum_{\mathrm{ALL} \ n} y^{kl}_n \chi_n(\vec{r}_1) \quad ,
\end{equation}
\textit{with an infinite sum over the product basis functions covering all space.}  In numerical calculations a finite basis is always used,
but by including all possible product basis functions in the above sum we avoid the need for a boundary condition term, 
as was needed (and easily handled) in the prior treatment in spherical polar coordinates \cite{topicalreview}.  
%
%
%
Applying the Laplacian with respect to $\vec{r}_1$ to this expansion gives
\begin{equation}
\label{eq:poissonexpand}
\nabla_{\vec{r}_1}^2 y^{kl}(\vec{r}_1) \approx \sum_{\mathrm{ALL} \ n}  y^{kl}_n \nabla_{\vec{r}_1}^2 \chi_n(\vec{r}_1).
\end{equation}
Multiplying Eqs.(\ref{eq:poisson}) and (\ref{eq:poissonexpand}) through by 
$\chi_m(\vec{r}_1)$, integrating over $\vec{r}_1$ and equating the results leads to
\begin{equation}
\sum_{\mathrm{ALL} \ n} y^{kl}_n T_{mn} = 2\pi \int d^3\vec{r}   \chi_{k}(\vec{r}) \chi_l(\vec{r}) \chi_m(\vec{r})
\label{reseq}
\end{equation}
where $T_{mn} = -\frac{1}{2}\langle \chi_m | \nabla^2 | \chi_n \rangle$ are the kinetic energy matrix elements. 

The integral on the right hand side of Eq.(\ref{reseq}) is evaluated by 
a resolution of the identity,  Eq.~\ref{interp},
such that the density (a sum of squares of localized basis functions) is 
 re-expanded as a sum of localized basis functions, without the square.
From Eqs.(\ref{reseq}) and (\ref{interp}),
\begin{equation}
\sum_{\mathrm{ALL} \ n}  y^{kl}_n T_{mn} = 2\pi  \delta_{kl}\delta_{lm} (w_k w_l)^{-1/2}
\end{equation}
giving
\begin{equation}
y^{kl}_n = 2\pi  \delta_{kl} (w_k)^{-1} (T^{-1})_{nk} \quad ,
\label{ydef}
\end{equation}
\textit{where $T^{-1}$ is the limit of the matrix inverse as the size of the matrix goes to infinity.}
Using these coefficients and inserting Eq.(\ref{yklexpand}) into Eq.(\ref{eq:twoelectronsimple}) 
gives 
\begin{equation}
\label{eq:twoelectronexpand}
[ij | kl] = \sum_n y^{kl}_n \int d^3\vec{r}_1  \ \chi_i(\vec{r}_1) \chi_j(\vec{r}_1)  \chi_n(\vec{r}_1)  \quad .
\end{equation}
Once again, the integral on the right hand side is evaluated by resolving the identity and 
employing DVR quadrature to obtain the final expression for the two electron matrix elements
\begin{equation}
[ij | kl] = 2\pi \delta_{ij} \delta_{kl} (w_i w_k)^{-1/2} T_{ik}^{-1} \quad .
\label{finalexprfull}
\end{equation}

\subsection{One-electron matrix element}

The expression for the one-electron matrix element 
%
%
for a nucleus at position $\vec{R}$, here denoted $U_{ij}$,
\begin{equation}
U_{ij}(\vec{R}) \equiv \left\langle i \left \vert \frac{1}{\vert \vec{r}-\vec{R}} \right \vert j \right\rangle 
\end{equation}
within the present method follows simply from the analog of
Eq.~\ref{yklexpand}, 
\begin{equation}
U_{ij}(\vec{R}) \approx \sum_{\mathrm{ALL} \ n} u^{ij}_n \chi_n(\vec{R}) \quad ;
\label{nucexpand}
\end{equation}
%
%
Eq.~\ref{interp}, which
is the resolution of the identity approximation; and Poisson's equation:
\begin{equation}
\nabla_{\vec{R}}^2     U_{ij}(\vec{R})  = 4\pi \chi_i(\vec{R}) \chi_j(\vec{R}) \approx 4\pi \Delta^{-3/2} \delta_{ij} \chi_i(\vec{R})
\label{oneelec}
\end{equation}
within the weak variational formulation, i.e., multiplying by the left by $\int d^3R \ \chi_k(\vec{R})$ for all $k$.
The one-electron matrix element is thereby simply related to the two-electron matrix element,
\begin{equation}
U_{ij}(\vec{R})  \approx y^{ij}(\vec{R}) \quad ;
\label{finalone}
\end{equation}
both are defined via Eqs.~\ref{yklexpand} and~\ref{ydef} in terms of the kinetic energy matrix elements.  However,
we find that accurate results are only obtained when the nuclei are at positions $\vec{R}$ coinciding
with electronic DVR grid points $\vec{r}_i$.  
In other words, for the moment, the method requires that the nuclei be placed on the Cartesian grid points,
which is a major limitation.  We do not understand this behavior and comment upon it, and ways to avoid it,
in the conclusion.  

In order to use the above expressions in a practical way, for general DVR bases, 
we must have a method of 
approximating the matrix elements of $T^{-1}$. 
 However, 
simply inverting the finite-dimensional kinetic energy matrix is not a good approximation and 
would also require the storage of a dense matrix, 
which is impractical for even modestly sized grids. 
We use the method below, which directly calculates the entire set of diagonal
two-electron matrix elements at one time, and which
would be applicable to other generalized DVR basis sets besides the Cartesian product 
sinc functions used here.

\subsection{Kinetic Energy Inverse \label{kesect}}

The key to a practical implementation is to consider the representation of $T^{-1}$ on an 
infinite grid, and then truncate the matrix elements to a finite grid. 
The full kinetic energy in three dimensions is given in Eq.~\ref{asusual}.
Since $t$ is Toeplitz, $T$ is Toeplitz with respect to each of the 
spatial indices, i.e. $T$ is triple Toeplitz, and therefore it could be denoted using a symbol with only
one three-vector index, e.g. $T_{\vec{i}\vec{j}} = u_{\vec{i}-\vec{j}}$.  (In this subsection we momentarily
revert to explicit vector index notation.) Its inverse inherits this property, 
\begin{equation}
(T^{-1})_{\vec{i}\vec{j}} = v_{\vec{i}-\vec{j}}
\label{inherits}
\end{equation}
The expression $T T^{-1}=1$ can be rewritten
\begin{equation}
\left( u*v \right)_{\vec{j}} \equiv \sum_{\vec{i}=-\infty,\infty,\infty}^{\infty,\infty,\infty} u_{\vec{i}-\vec{j}} v_{\vec{i}} = \delta_{j1,0}\delta_{j2,0}\delta_{j3,0}
\label{convolve}
\end{equation} 
where $*$ represents the discrete convolution product.

For the sinc DVR, we may derive an exact expression for the matrix element amenable to quadrature, but instead
we solve for all of the matrix elements simultaneously
using the following method, which is also applicable to other bases.  
The strategy is to take the lowest-order Taylor series expression for the matrix element,
for those matrix elements at long range, and solve for the remainder.

Thus we approximate the matrix elements of $T^{-1}$ between basis functions far separated as
\begin{equation}
v_{\vec{i}} \stackrel{|i|\rightarrow\infty}{\longrightarrow} \frac{\Delta^3}{2\pi} \frac{1}{r_{\vec{i}}} = \frac{\Delta^2}{2\pi} \frac{1}{\sqrt{i_1^2+i_2^2+i_3^2}}
\label{assume}
\end{equation}
such that $[ij | kl] \longrightarrow \delta_{ij}\delta_{kl}\frac{1}{r_{i-k}}$.  
We assume that Eq.~(\ref{assume}) holds exactly for three-indices $\vec{i}$ in which $i_1$, $i_2$, or $i_3$ is greater than $n_{small}$,
where $n_{small}$ is an adjustable parameter, and solve Eq.~(\ref{convolve}) for the remainder of $v$.  In other words, we solve
\begin{equation}
\sum_{\vec{i}=-n_{big},n_{big},n_{big}}^{n_{big},n_{big},n_{big}} u_{\vec{i}-\vec{j}} v_{\vec{i}}  = \delta_{j1,0}\delta_{j2,0}\delta_{j3,0} \\
\label{convolve2}
\end{equation}
given $u_{\vec{i}-\vec{j}}=T_{\vec{i}\vec{j}}$ and
\begin{equation}
v_{\vec{i}} = \frac{\Delta^2}{2\pi} \frac{1}{\sqrt{i_1^2+i_2^2+i_3^2}} \qquad \left(
\begin{array}{l}
| i_1 | >n_{small}  \ \mathrm{or}   \\
\ | i_2 | >n_{small}  \ \mathrm{or} \\
 \ | i_3 | > n_{small} 
\end{array} 
\right)
\end{equation}
for the remaining $(2n_{small}+1)^3$ elements of $v$.  
The infinite sum in Eq.~\ref{convolve} is truncated at $n_{big}$ and we therefore
have two convergence parameters, $n_{small}$ and $n_{big}$ defining the approximated $T^{-1}$.
We have chosen a default of 40 and 240 for these numbers, respectively, and we 
verify the convergence as a function of these parameters of all the results presented below.

\subsection{Triple Toeplitz linear algebra with Fourier transforms\label{toepsect}}

We continue with vector index notation in this section, after which we revert to condensed index notation.
 
To construct a two-electron matrix element among contracted basis functions
\begin{equation}
\phi_\alpha(\vec{r}) = \sum_{\vec{i}} c_{\vec{i}\alpha} \chi_{\vec{i}}(\vec{r})
\end{equation}
we must perform the sum
\begin{equation}
[\alpha\beta\vert\gamma\delta] = \sum_{\vec{i}\vec{k}} 2\pi (w_{\vec{i}}w_{\vec{k}})^{-1/2} v_{\vec{i}-\vec{k}} c_{\vec{i}\alpha}^* c_{\vec{i}\beta} c_{\vec{k}\gamma}^* c_{\vec{k}\delta}
\end{equation}
wherein a triple Toeplitz matrix-vector multiplication is performed by the triple Toeplitz matrix $v$ upon the density $\phi_\alpha^*\phi_\beta$ to produce a potential that is then integrated over the density $\phi_\gamma^*\phi_\delta$, or vice versa.

The matrix $T^{-1}$ is triple Toeplitz (a.k.a., 3-level Toeplitz), i.e., 
\((T^{-1})_{ijk,i'j'k'}=v_{i-i',j-j',k-k'}\) where \(v\) is a \((2l-1)\times (2m-1)\times (2n-1)\) tensor 
and \(T^{-1}\) is an \(N\times N\) matrix with \(N=lmn\). 
A triple Toeplitz matrix is (a) block Toeplitz, e.g.,
\[T^{-1}=\begin{bmatrix}     (T^{-1})_0 &      (T^{-1})_1 & \cdots &     (T^{-1})_n \\
                     (T^{-1})_{-1} &      (T^{-1})_0 &     \cdots & (T^{-1})_{n-1} \\
                     \vdots &   (T^{-1})_{-1} & \ddots &  \vdots \\
                   (T^{-1})_{-n-1} &   \vdots & \ddots &      (T^{-1})_1 \\
                     (T^{-1})_{-n} & (T^{-1})_{-n-1} & \cdots &      (T^{-1})_0

    \end{bmatrix}\]
Furthermore, (b) the blocks are double Toeplitz (or 2-level Toeplitz), i.e., they are block 
Toeplitz with Toeplitz blocks (also called BTTB in the literature).
Similarly,
a triple circulant matrix \(C\) is such that 
\(C_{ijk,i'j'k'}=c_{i-i'\pmod{l},j-j'\pmod{m},k-k'\pmod{n}}\) 
(\(c\) is an \(l\times m\times n\) tensor, \(C\) is an \(N\times N\) matrix, \(N=lmn\)).

Triple circulant matrices are diagonalized by the three-dimensional Fourier transform 
(Theorem 5.8.4 in~\cite{davis1979circulant}):
    \[C=F'\cdot \mbox{diag}(Fc)\cdot F\]
An \(N \times N\) triple Toeplitz matrix
can be embedded into an \(8N \times 8N\) triple circulant matrix~\cite{chen2013parallelizing}.
Therefore, just as with single Toeplitz~\cite{briefsurvey}, 
a matrix-vector product involving a triple Toeplitz matrix, such as that required to
compute two-electron matrix elements,
 may be computed in \(O(N\log N)\) floating point operations 
using a fast Fourier transform, instead of \(O(N^2)\).
Memory use is minimal, due to the redundancy inherent in Toeplitz matrices.

The embedding that is used~\cite{chen2013parallelizing}
to transform the \(N\times N\) triple Toeplitz matrix into an 
\(8N\times 8N\) triple circulant matrix consists 
in padding the tensor \(v\) with zeros.  We summarize the algorithm:

\begin{equation*}\small
\begin{array}{|l|}
\hline
\mbox{Matrix-vector product } y=(T^{-1})x \mbox{ with }                               \\ 
T^{-1}: N\times N\mbox{ \textbf{Triple Toeplitz matrix}, } N=lmn, \mbox{defined by} \\ 
v: (2l-1)\times(2m-1)\times(2n-1) \mbox{ tensor per Eq.~\ref{inherits}}                      \\ 
\hline
1\!:\ x_2=0\hfill (2l\times2m\times2n)                                         \\ 
2\!:\ x_2(1:l,1:m,1:n)=x(:,:,:)                               \\ 
3\!:\ v_2=0\hfill (2l\times2m\times2n)                                         \\ 
4\!:\ v_2(2:2l,2:2m,2:2n)=v(:,:,:)                                             \\ 
5\!:\ f_x=\textrm{FFT-3D}(x_2)                                                 \\ 
6\!:\ f_v=\textrm{FFT-3D}(v_2)                                                 \\ 
7\!:\ f_y=f_x\times f_v\hfill\mbox{(element-wise product)}                     \\ 
8\!:\ y_2=\textrm{IFFT-3D}(f_y)                                                \\ 
9\!:\ y(:,:,:)=y_2(l+1:2l,m+1:2m,n+1:2n)                             \\ 
\hline
\end{array}
\end{equation*}

\section{Results} \label{resultsect}

\begin{center}
\begin{table}
\begin{tabular}{|c|c|c|c|c|c|c|c|}
\hline
& \multicolumn{3}{|c|}{Energy} & \multicolumn{3}{|c|}{Virial theorem} \\
\hline
State          &  DVR             & Variational      & Exact &    DVR        & Var.  & Ex. \\
\hline
1$s$           & $-1.9765$     &   $-1.9526$   &-2 & $-0.4939$   & $-0.4817$    &-0.5    \\
2$p$          & $-0.4998$      &   $-0.4953$   &-0.5 & $-0.4998$   &  $-0.5184$    &-0.5 \\
2$s$          &  $-0.4976$     &   $-0.4826$   &-0.5 & $-0.4987$   & $-0.5280$  &-0.5\\
3$d^\dag$ &  $-0.2189$     &   $-0.1761$   &-0.22... & $-0.5282$   &  $-0.7108$ &-0.5\\
3$d^*$       & $-0.2155$     &   $-0.1712$    &-0.22... & $-0.5494$   & $-0.7184$ &-0.5 \\
\hline
 \end{tabular}
\caption{The $n=1,2,3$ eigenvalues and virial theorem ratios $<T>/<V>$ for He$^+$ using both the 
DVR and the variational method with grid spacing $\Delta=0.4a_0$. 
{\scriptsize
$^\dag$ $xy$, $yz$ and $xz$ components    
\ $^*$ $2z^2$ - $x^2$ - $y^2$ and $x^2-y^2$ components } 
\label{table:eigs}}
\end{table}
\end{center}

\subsection{DVR Method vs. Variational Method}

Remarkably, the treatment we have outlined appears to perform better than the variational method 
(in which the Coulomb matrix elements are evaluated exactly).  We have not
been able to test the variational method for a two-electron problem with the methods available to us, 
due to the prohibitive cost of computing and storing matrix elements.  Here we compare the results for 
the hydrogen atom, which tests the one-electron operator.

In Table \ref{table:eigs}, we show the  $n=1,2,3$ eigenvalues of He$^+$ and the ratios 
$\langle T \rangle / \langle V \rangle$ for both the DVR and variational methods, using a grid spacing of $\Delta=0.4a_0$.
The present DVR method clearly gives better energies, in one case (2p) by nearly two orders of magnitude.  Results for the virial theorem
are even more decisive, up to three orders of magnitude.  
We speculate that the reason for this favorable performance is that the
relationship between the Coulomb potential and the kinetic energy operator
is maintained in matrix form.

\begin{figure}
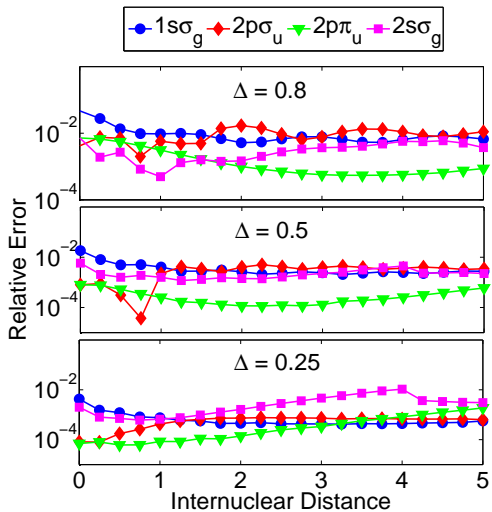

\begin{center}
\begin{tabular}{c}
\resizebox{0.8\columnwidth}{!}{\includegraphics{{{h2pluserror3}}}}
\end{tabular}
\end{center}
\caption{The relative error in the $n=1,2$ energies of H$_2^+$ vs. internuclear distance, 
with $\Delta=$0.25, 0.5, and 0.8$a_0$, from $R = 0$ to $5a_0$.
\label{fig:h2pluserror}}
\end{figure}

\subsection{H$_2^+$}

In Figure~\ref{fig:h2pluserror} we show the relative error in the $n=1,2$ energies of H$_2^+$ for different grid resolutions, 
compared with exact results obtained in prolate spheroidal coordinates as in Ref. \cite{tao+mccurdy}.
The errors are on the order of a millihartree for $\Delta=$ 0.5 and 0.8$a_0$ and are 1-2 orders of magnitude better with $\Delta=0.25a_0$.
It appears that the $2p\pi_{\textnormal{u}}$ state is generally the most accurate. 
The errors are also relatively constant with respect to the internuclear distance.

\begin{figure}
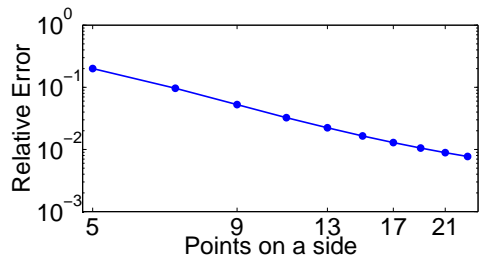

\begin{center}
\begin{tabular}{c}
\resizebox{0.8\columnwidth}{!}{\includegraphics{{{helium_convergence}}}}
\end{tabular}
\end{center}
\caption{The relative error in the ground state energy of He vs. the number of grid 
points per spatial dimension with a fixed box size of 3$a_0$, plotted on a log-log scale. 
\label{fig:heliumconvergence}}
\end{figure}

\subsection{Two-electron results}

\begin{figure}
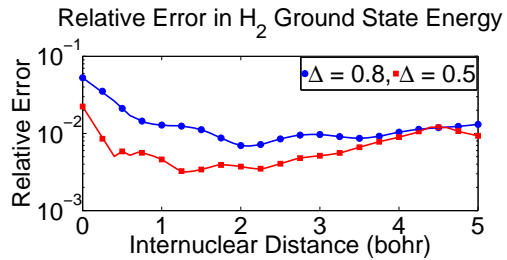

\begin{center}
\begin{tabular}{c}
\resizebox{0.8\columnwidth}{!}{\includegraphics{{{h2error}}}}
\end{tabular}
\end{center}
\caption{The relative error in the ground state energy of H$_2$ vs. internuclear distance, 
with $\Delta=$0.5 and 0.8$a_0$, from $R = 0$ to $5a_0$,
defined relative to the benchmark results of Ref. \cite{sims+hagstrom}.
\label{fig:h2error}}
\end{figure}

In Figure~\ref{fig:heliumconvergence}, we plot the relative error in the ground state of Helium for multiple grid
resolutions with a fixed box size of $3a_0$, which is sufficient to eliminate truncation error. 
The figure demonstrates a roughly quadratic convergence rate of the ground state 
energy of Helium with respect to the grid resolution; the error is proportional to $N^{p}$ where $p \approx -2.13$.

In Figure~\ref{fig:h2error} we show the  relative error in the ground state energy of H$_2$ for different grid resolutions.  
As expected, the results with $\Delta=0.5a_0$ are more accurate than with $\Delta=0.8a_0$.
However, both resolutions have error roughly on the order of $10^{-2}$, with the errors being slightly larger for $R$ close to zero. 
Comparing the ground state errors of H$_2$ with those of H$_2^+$, we see that the H$_2$ calculations are slightly 
less accurate, due to the error introduced by the two-electron operator.

\begin{figure}
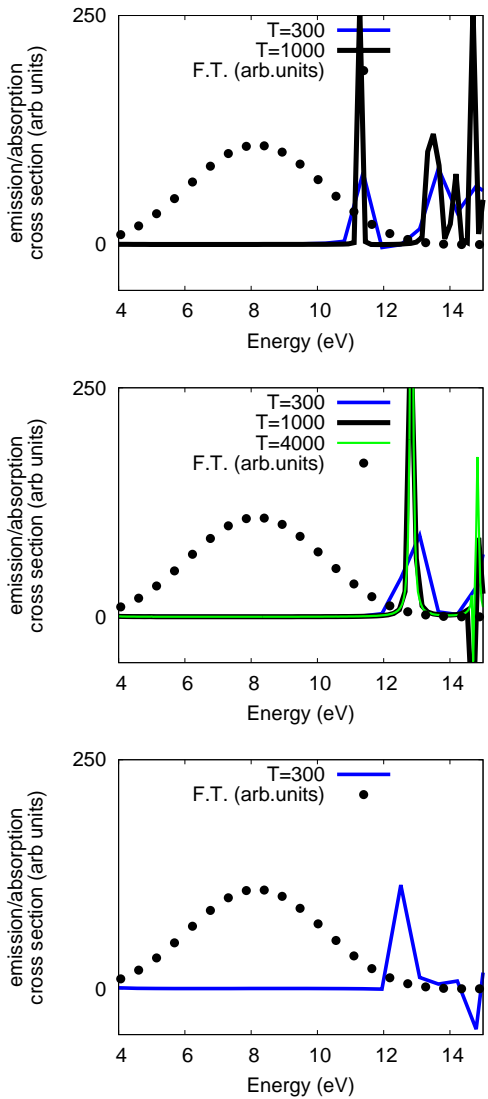

\begin{center}
\begin{tabular}{c}
\resizebox{0.8\columnwidth}{!}{\includegraphics{{{Absorption.0.40186.bigger2}}}} \\
\resizebox{0.8\columnwidth}{!}{\includegraphics{{{Absorption.0.40186.bigger}}}} \\
\resizebox{0.8\columnwidth}{!}{\includegraphics{{{Absorption.0.20093.bigger}}}} 
\end{tabular}
\end{center}
\caption{Bound-state absorption spectrum for methane calculated 
one pulse and equal wall clock computation time.
Top to bottom:
Grid spacing 0.39495414$a_0$, 63 points on a side; grid spacing 0.39495414$a_0$, 31 points on a side; 
grid spacing 0.19747707$a_0$, 63 points on a side.
Time is in atomic units; divide by 41.34 for femtoseconds.\label{methabs}}
\end{figure}

\subsection{Methane photoexcitation, time-dependent calculation}

We calculate time-dependent nine-orbital full configuration interaction electronic wave functions for methane
using the method described in Ref.~\cite{mctdhfpaper}.   We use grid spacings $\Delta$ of
0.39495414 and 0.19747707 bohr.  These spacings permit a bond length of  of 1.086\AA, which is the
equilibrium bond length at the highest level of theory in Ref.~\cite{harr_meth}.  We position hydrogen nuclei at
$(x,y,z)=\{(A,A,A), (A,-A,-A), (-A,A,-A), (-A,-A,A)\}$ where $A=3\Delta$ or $A=6\Delta$, respectively, for the two
grid resolutions.
The absorption spectra are calculated in a straightforward manner, as in Ref.~\cite{mynopaper}, and shown
in Fig.~\ref{methabs}.

We perform three nine-orbital full-configuration-interaction calculations --
15876 Slater determinants, 5292 spin adapted singlet configurations --
 with different grid bases but otherwise
identical, and approximately equal total wall 
clock computation time.  Due to the different rates of calculation, different final times are reached:
\begin{itemize}
\item{Resolution $\Delta$= 0.39495414$a_0$, 63 points on a side: $t_{final}\approx$7.5fs (300 atomic units), or}
\item{Resolution $\Delta$= 0.39495414$a_0$, 31 points on a side: $t_{final}\approx$24.2fs (1000 atomic units), or}
\item{Resolution $\Delta$= 0.19747707$a_0$, 63 points on a side: $t_{final}\approx$121fs (5000 atomic units)}
\end{itemize}
The difference in speed between the two 63-point calculations is due to the different behavior of the method for the different resolution grids.
The difference between the two calculations with coarse resolution is due to the size of the basis.  The basis size affects the Fourier transform
time (O($n^3 \log(n)$)) and the kinetic energy time (O($n^4$), but with a smaller coefficient).  

Excitation of the low-lying valence states leads to photodissociation; photodissociation of methane
has been studied by several authors~\cite{mebel_chang, harr_meth, harr_meth2, schatz_meth}.
The vertical excitation energy of the lowest state was calculated in Ref.~\cite{harr_meth} to be 10.60eV and this appears
to be the most reliable value in the literature.  We have used the same bond length as in that work.

One can see that, of the three panels shown in Fig.~\ref{methabs}, the first one shows a peak that is just above 11eV,
whereas the peak in the other panels occurs higher, between 12 and 13eV.  The difference 
is due to the truncation error, the error caused by insufficient spatial extent of the grid; 
the grid used for the top panel is about twice as wide as that used for the other two.
The difference between the middle and bottom panels in this figure is due to resolution.  The grids have approximately
the same extent in these figures, but the bottom figure uses a grid spacing $\Delta\approx$0.2$a_0$, whereas
the middle uses $\Delta\approx$0.4$a_0$.  The difference in the $t=300$ curves on the middle and bottom panels is minor.
So we see that a box size of 24$a_0$ and a grid spacing of 0.4$a_0$ are probably sufficient for calculations
of excited state physics on methane.

The fact that we must use a large basis set to represent electronic wave functions with large spatial extent is certainly
a problem with the representation, and we comment on the possibility of stretching the grid in the asymptotic region later,
in the conclusion.  We would like to avoid truncation error, what comes from a grid of insufficient spatial extent, in
the analysis of this basis set method, and focus on resolution error.
In subsections~\ref{methexcite} and~\ref{cubanesect}, we 
perform apples-to-apples comparisons of the sinc DVR to Gaussian basis sets, in which
we eliminate truncation error, which does not apply to Gaussian basis sets, from consideration.
The method we use for this purpose is described below.

\subsection{Method for extrapolation to infinite grid size\label{extrapsect}}

Below we report transition energies and ionization potentials for polyatomic molecules.
The desire is to present the results without error due to the use of a grid with finite extent.
The ``truncation error,'' the error due to having an insufficient number
of points on a side $n$, is uninteresting and should be eliminated.

The resolution error, in contrast, is of prime importance.  We have conjectured that this is
an ideal smoothed Coulomb representation on a Cartesian grid.  
So we report numbers that are functions of resolution $\Delta$, but not of points on a side $n$.
However, they have error bars due to
the method used to extrapolate them, as a function of points on a side $n$, to $n=\infty$.
We report excitation energies as a function of resolution $\Delta$, with error bars due to truncation error
in our finite basis calculations.
The method we use to extrapolate the energies is ad hoc and is as follows.  

We choose a function $f(n; \vec{P})$ that is monotonic as a function of $n$ and that approaches a limit as $n\longrightarrow \infty$ but that is otherwise
arbitrary, and that is a function not only of $n$ but also of certain number $N_P$ of parameters $P_i$, $i=1...N_P$.
Each eigenvalue $E_\alpha(n)$, $\alpha=1...12$ is fit separately as a function of $n$ to the function $f$ by varying the parameters of $f$.

The uncertainty in the extrapolated energy eigenvalue will be affected by the choice of $f$.  We regard $f$ as unknown and seek to find
one that provides acceptable precision in the reported extrapolated eigenvalue.  Presently we have tried functions of
the form $P_1 + P_2 n^Q e^{-n P_3}$ and find that 
\begin{equation}
f(n; P_1, P_2, P_3) = P_1 + P_2 n^2 e^{-n P_3}
\label{expon}
\end{equation}
gives a consistently superior fit to the present data, when compared to the other choices we tried with $Q\ne 2$, 
so we chose this function, with an $n^2$ factor in the exponential term, for $f$.
We perform a least squares regression, choosing $N_C$ values of $n$ with which to perform calculations and
minimizing
\begin{equation}
\forall_{\alpha=1...12, i=1...N_P} \ \frac{\partial}{\partial P_{\alpha i}} \sum_{j=1}^{N_C}
\left[ E_\alpha(n_j) - f(n_j; \vec{P}_\alpha) \right]^2 = 0
\end{equation}
 The predicted asymptote
is the first parameter, $P_1$ from equation~\ref{expon}, the constant term,
\begin{equation}
E_\alpha(\infty) \equiv P_1
\end{equation}

The variance in the predicted $E_\alpha(\infty)$ will be denoted $\sigma^2_\alpha$.
There is systematic error in the prediction due to the lack of knowledge about the exact form of the unknown function $f$.  There
is statistical error due to imperfect convergence of the calculated eigenvalues $E_\alpha(n)$.  So we estimate
the variance as
\begin{equation}
\sigma_\alpha^2 = \left(\sigma_\alpha^{sys}\right)^2 + \left(\sigma_\alpha^{stat}\right)^2
\end{equation}
with the systematic error defined as the asymptotic standard error of the parameter $P_1$.

The statistical error for each computed 
eigenvalue $E_\alpha(n)$ is that caused by imperfect convergence of the MCTDHF relaxation procedure.
 We have a primitive implementation but choose a stringent convergence criterion.
The change in energy between the penultimate and final iterations, which we will denote $\Delta E_\alpha(n)$, 
is generally less than one microhartree, 
and this number is recorded for each eigenvalue and used to estimate
$\sigma_\alpha^{stat}$.  We performed several small runs with an error criterion even more stringent.
We estimate that the change in energy between the penultimate and final iterations is significantly more
than 100 times the error in the final eigenvalue, and therefore we conservatively estimate the
statistical error for each point separately as
\begin{equation}
\sigma_\alpha^{stat}(n) \equiv 100 \times \Delta E_\alpha(n) \quad .
\end{equation}
Given that 
these individual statistical errors may be correlated, we define the statistical error of the overall fit as the average of them,
\begin{equation}
\sigma_\alpha^{stat} \equiv \frac{1}{N_C} \sum_{j=1}^{N_C} \sigma_\alpha^{stat}(n_j)
\end{equation}

In summary, we conservatively define the variance in the fitted asymptote, the variance in 
fitted value of the transition energy in the limit of infinite basis size, as
\begin{equation}
\sigma_\alpha^2 \equiv \left(\sigma_\alpha^{sys}\right)^2 + \left(\frac{100}{N_C} \sum_{j=1}^{N_C} \Delta E_\alpha(n_j)\right)^2
\end{equation}

Furthermore, we perform two calculations with different values of the parameters $n_{big}$ and $n_{small}$ in order to check the
error due to the approximations made in our calculation of the Coulomb matrix elements.  The significant figures
reported in sections~\ref{methexcite}
and~\ref{cubanesect} agree for the two choices $(n_{big},n_{small})$ = (248,31) and (195,39).

\subsection{Methane excitation energies\label{methexcite}}

We calculate excitation energies of methane using the same nine-orbital full-configuration-interaction representation
used for the time-dependent calculations above, using two
The calculation we perform is called state-averaged
multiconfiguration self-consistent field (MCSCF) and consists of minimizing the average energy of the first twelve
electronic states of methane with respect to variations both of the coefficients of the sinc DVR basis functions comprising
the nine orbitals, and of the coefficients of the spin-adapted linear combinations of Slater determinants.

These energies are calculated as a function of grid resolution, independent of box size (points on a side $n$), but
with error bars that are due to finite box size calculations, using the method described in the subsection immediately above,
and reported in Table~\ref{methtable}.  Nine or eight calculations are used for the extrapolation to infinite
basis size, respectively: for $\Delta$=0.39495414$a_0$, 
 $n=$105, 115, . . . 185; for $\Delta$=0.19747707, $n=$135, 145, 155, . . . 215.

For comparison, 
we perform the same state-averaged MCSCF calculations using Gaussian basis sets, using the Columbus
suite of codes for quantum chemistry~\cite{columbus}.  We use three basis sets,
aug-cc-pvdz, aug-cc-pvtz, and aug-cc-pvqz~\cite{dunning}, using either the full set of Cartesian
basis functions or contracting them to make spherical harmonics.  These results are reported in Table~\ref{gausstable}.

There are only two columns in Table~\ref{methtable}, for only one molecule; any conclusions about the method at this
stage must be considered preliminary.  
The columns in Table~\ref{methtable}, the results with the current sinc basis, differ consistently
by about 0.1eV.  The double-zeta and triple-zeta columns in Table~\ref{gausstable}, obtained with standard Gaussian basis
set methods, have a range of differences, from 0.05 to 0.16eV.  
Therefore, it appears that rougly double-zeta accuracy is obtained with a grid spacing 
 $\Delta$=0.39495414$a_0$, and roughly triple-zeta accuracy is obtained
with $\Delta$=0.19747707.  We perform a more quantitative analysis of the performance of the representation
as a function of grid resolution $\Delta$ in the next section, on cubane.

\begin{table}
\begin{tabular}{l lll}
$\Delta=$ &  0.39495414$a_0$  & 0.19747707 \\
\hline
T & 9.666150(1)eV  & 9.5353(5)  \\
E & 10.718654(3) & 10.6309(2)  \\
T & 10.772476(3) & 10.6826(2)  \\
T &10.773745(3) & 10.6850(2)  \\
\end{tabular}
\caption{Transition energies $E_\alpha(\infty)$ for methane calculated with 12-state-averaged MCSCF using the sinc DVR basis, 
for two grid resolutions, calculated by extrapolating to infinite basis size
using the method of Sec.~\ref{extrapsect}. \label{methtable}}
\end{table}

\subsection{Cubane (C$_8$H$_8$) ionization potential \label{cubanesect}}

As presently described, without elaboration,
this representation for electronic wave functions of molecules using the sinc discrete variable representation (DVR)
requires that nuclei be placed on the Cartesian grid points and as such, has limited applicability.
 In the conclusion we speculate about elaborations to the method that would allow
it to calculate a molecule in an arbitrary internuclear geometry.

%
%
%
%
%
%

For the moment, the cubane molecule
provides a good test of the method due to its cubic geometry.  Not only is it cubic, but the C-C and H-H
distances are approximately in the ratio 9:5 or 1.8.  The theoretical equilibrium geometry calculated at the
coupled cluster with single and double excitations (CCSD) using the cc-pVDZ Dunning basis set~\cite{dunning},
as tabulated by NIST~\cite{cccbdb},  has the carbons at $x, y, z = \pm 0.7893$ Angstrom and the
hydrogens at 1.4248 Angstrom, a ratio of 1.805.  So we take the geometric average of these distances, and multiply
and divide by the square root of 1.8, to arrive at our geometry, with the carbons at $x, y, z = \pm 1.493691a_0$ 
(approximately 0.7904 Angstrom) and the hydrogens at $x, y, z = \pm 2.6886438a_0$ (approximately 1.4228
Angstrom).  We use three grid resolutions, $\Delta =$ 0.5974764, 0.2987382, and 0.1493691$a_0$.

\begin{table}[!t]
\begin{tabular}{l|cccccc}
    & DZ-s             & DZ            & TZ-s                 & TZ                & QZ-s          & QZ  \\
T  & 9.5845    & 9.5828      & 9.5381        &   9.5359   &      9.5197  &     9.5160 \\   
T  & 10.8820  & 10.8815   & 10.7792      & 10.7774  &     10.7462 &    10.7419 \\  
E  & 10.9799  & 10.9795  & 10.8225      & 10.8173   &     10.7585 &    10.7517 \\  
T  & 11.0339  & 11.0335    & 10.8792     & 10.8740   &    10.8157 &    10.8089 \\  
\end{tabular}
\caption{12-state-averaged MCSCF energies calculated with Gaussian basis sets,
aug-cc-pvdz, aug-cc-pvtz, and aug-cc-pvqz, either the full Cartesian basis (no extension) 
or contracting them spherically (extension -s above).
\label{gausstable}}
\end{table}

In Table~\ref{cubetable}, we present results showing the first two ionization potentials of cubane in the Hartree-Fock approximation, 
calculated as in Sec.~\ref{extrapsect}, 
extrapolated to infinite basis size, for the three grid spacings $\Delta$.
These infinite-basis results are then extrapolated to $\Delta=0$, and that result is shown in the fourth row of the
table.  The method that we use for this final extrapolation is described later in this section.

Two potentials are reported, both those corresponding to the difference between the fully converged Hartree-Fock
energies of the neutral and cation, labeled ``I.P.'' in Table~\ref{cubetable}, and those corresponding to the Koopman's 
ionization potential, labeled ``K.I.P.,'' corresponding to the neutral Hartree-Fock highest occupied molecular orbital energies, calculated
as the difference between the neutral Hartree-Fock energy and the cation energy obtained through diagonalization
using the neutral Hartree-Fock orbitals.  The precision obtained in the latter is much lower than the former due
to the primitive Hartree-Fock implementation we use.  Five or six points are used for the extrapolation to infinite
basis size; for resolution $\Delta=0.5974764a_0$, $n=$64, 72, 80, 90, 108;
for $\Delta=0.2987382$, $n=$81, 91, 99, 105, 117; and for $\Delta=0.1493691a_0$, $n=$185, 195, 205, 215, 225, and 235.

  However, the precision
in the results for cubane in Table~\ref{cubetable} does not come from the extrapolation.  For the ionization potentials
(I.P.) the precision
comes from disagreement between the two calculations for
the different choices for $n_{big}$ and $n_{small}$; for the Koopman's ionization potentials (K.I.P.) the precision
comes from nonconvergence of the primitive Hartree-Fock procedure, and our conservative choice for the definition
of statistical error based upon it.

The ionization potentials of cubane have been previously calculated in 
Refs.~\cite{theoretical_cubane, spectroscopic_cubane, vertical_cubane}.  The lowest $T_{2g}$ and $T_{2u}$ 
Koopmans' ionization
potentials, exactly analogous to those reported here, were calculated to be 10.39 and 10.58eV at the double zeta with polarization
level of theory, in Ref.~\cite{vertical_cubane}.  In a different basis, the K.I.P.s were 10.42 and 10.59eV, and the
delta-SCF result, closely comparable to the I.P. reported here, was 9.74eV for both the $T_{2g}$ and $T_{2u}$ states.
 In Ref.~\cite{spectroscopic_cubane}, the Koopmans' I.P.s were
calculated as 10.40 and 10.62eV, respectively, and the I.P.s were calculated
to be 9.38 and 9.73eV at a higher level of theory with more correlation.

By comparing these numbers from the literature to those in Table~\ref{cubetable}, it seems that the present representation
will be able to produce qualitatively accurate results on polyatomic molecules using a grid spacing of approximately 0.3$a_0$,
the medium resolution in the table.  At this medium resolution, it seems that double-zeta quality transition energies are obtained;
stepping up to the finest resolution produces improvements of less than 0.1eV for the lower cation ionization potential, 
and improvements slightly greater for the higher I.P.  This accuracy of 0.1eV is unsatisfactory for many applications involving ground-state
Born-Oppenheimer dynamics; the standard called for there, ``chemical accuracy,'' is one kilocalorie per mole~\cite{chemical_accuracy}, 
which is approximately 43meV.  
Examining the lowest-resolution results, one can see
that errors introduced going from the resolution of approximately
$\Delta=0.3a_0$ to 0.6$a_0$ are a substantial fraction of an electron volt.
The accuracy at 0.6$a_0$ is probably unsatisfactory for almost all applications, but
$\Delta=0.3a_0$ seems sufficient
for qualitative studies of excited state potential energy curves and time-dependent electron 
dynamics of polyatomic molecules.

We have attempted to quantify the performance more accurately by extrapolating the results in the first three rows of
Table~\ref{cubetable} to $\Delta=0$.  The power law for the error that was observed for the one electron results and reported
above -- a power law $\Delta^{2.13}$ for the error -- does not fit the results on cubane in Table~\ref{cubetable}.
Unfortunately, the exponent in the power law for these cubane results is significantly lower.
In order to obtain error bars on the predicted extrapolation, we fit the four columns of Table~\ref{cubetable}
to the same power law.  In other words we consider the columns in the table to be labeled $E_i(\Delta)$, and with these
twelve points fit the 
nine parameters $\{e_1 . . . e_4, b_1 . . . b_4, Q\}$ in the functional form
\begin{equation}
E_i(\Delta) = e_i + b_i \Delta^Q
\end{equation}
using this fit we obtain the power law exponent $Q=1.205 \pm 0.18$.
The error bars in the final row of Table~\ref{cubetable},
showing this extrapolation, are almost entirely due to the error of this fit, and not to the error of the points
used in the fit.

Using the values of $b_i$ from this fit, we obtain the value at which the accuracy of the computed ionization potentials
for cubane is one kilocalorie per mole or 43meV, also known as ``chemical accuracy''~\cite{chemical_accuracy}.  By solving
$0.043 = b_i \Delta^{1.205}$, 
in electronvolts, for $\Delta$, 
given the fitted $b_i$, we obtain $\Delta=0.139a_0$  from both the 
I.P. and the K.I.P. of the lower 
($^2T_{2g}$) state, and approximately $\Delta=0.06a_0$ for both I.P. and K.I.P. for the upper ($^2T_{2u}$) state.  The average
of these values is about 0.1 bohr.  Given the flexibility of the representation, 
it seems reasonable to expect that chemical accuracy will be obtained generally, for other molecules as well, at this
resolution.
In the conclusion, we mention improvements to the method that would account for the truncation of the basis
in momentum space and that would hopefully yield chemical accuracy with an even lower resolution.


\section{Conclusion}

\begin{table}
\begin{tabular}{c|cc|cc}
& \multicolumn{2}{c|}{$^2$T$_{2g}$}  & \multicolumn{2}{c}{$^2$T$_{2u}$} \\
Resolution & I.P. & K.I.P. & I.P. & K.I.P. \\
\hline
0.5974764 & 9.84027(1)     & 10.68006(1) & 10.24536(2) & 11.16661(2) \\
0.2987382 & 9.69775(2)     & 10.54452(50)  & 9.84952(2) & 10.79933(40) \\
0.1493691 & 9.63852(10)    & 10.47702(10)  & 9.69365(4) & 10.64553(4) \\
0         & 9.591(2)       & 10.433(5)     & 9.560(9)   & 10.523(3) \\
\end{tabular}
\caption{Ionization potentials of cubane, in electronvolts, as described in the text.
\label{cubetable}}
\end{table}

We have demonstrated an efficient real-space basis set representation for electronic 
structure using sinc basis functions, a generalization of the 
method of Ref~\cite{topicalreview} to Cartesian coordinates.  This and that
method make use of a resolution-of-the-identity approximation to arrive
at diagonal expressions for the one- and two-electron matrix elements.  
The singular Coulomb potential is discarded and the one- and two-electron matrix elements
are obtained instead from the kinetic energy matrix elements
by requiring that the relationship between the Coulomb potential and Laplace
operator -- that the former is the Green's function of the latter -- 
be maintained in their numerical matrix representations.
The normally 
forth-rank tensor of two-electron matrix elements is rendered first-rank, and may
be stored in memory for even the largest problems.  
The energies and virial theorem ratios calculated are far superior to those obtained with
the variational method using the same sinc basis.

We note that this DVR representation bears similarity to that of Ref.~\cite{rob_extreme}, 
a three-dimensional treatment for atoms in spherical coordinates.  
The three dimensional representations, the present one and that of Ref.~\cite{rob_extreme}, 
as opposed to the treatment for spherical coordinates in Ref.~\cite{topicalreview}
in which only the radial coordinate is discretized, 
permit the maximum degree of parallel computer
scalability.  We also note that a similar \textit{ansatz} involving the kinetic energy operator
has been applied to Gaussian basis functions in Ref.~\cite{mhg}.

The representation described in this paper 
may provide a foundation for an efficient treatment of electronic 
structure that would compete with Gaussian basis set methods in applications
for which chemical accuracy~\cite{chemical_accuracy} is required. With this goal in mind, 
several easy-to-implement elaborations that would improve its performance are 
conceivable.  For instance, it is desirable to have grids with
different resolutions for different electrons in the Slater determinant basis, 
such that different orbitals with different spatial extents can be 
described efficiently.  Algebra along these lines is presented the Appendix.
Also, effective theory~\cite{lepage} may be used to account for the truncation in momentum space and improve the convergence
of the results with respect to resolution $\Delta$.

However, the most important improvement
is to permit small grid distortions.  Presently, the method requires that nuclei only be 
placed on the Cartesian grid points, which is its most
major limitation.  If the grid could be distorted slightly, but arbitrarily, then arbitrary 
internuclear geometries could be calculated simply by
distorting the grid such that the grid points and nuclei coincide.  Furthermore, if these 
distortions could be made complex-valued, then the representation
would be capable of calculating ionization using the method of complex 
coordinate scaling~\cite{abc1,abc2,moi2,moi3,ReinhardtReview82}.

Implementing complex-valued grid distortions is therefore the next step in the development of this real-space
representation for electronic structure.  Including grid distortions in the method
will permit arbitrary fixed-nuclei geometries and the accurate representation of ionization.   It will also enable 
calculations of 
fully nonadiabatic electronic and nuclear dynamics of polyatomic molecules subject to intense, ultrafast laser light,
with the open-source implementation published in Refs.~\cite{mctdhfpaper, restrictedpaper, lbnl_amo_mctdhf}.
Because of its efficiency and uniform resolution, 
  this sinc discrete variable representation for electronic structure 
is best suited to highly correlated, highly excited dynamics of electrons in molecules, not ground state
electronic structure.  With the method of Domcke and coworkers~\cite{wave1,wave2,wave3}, we are using it to calculate phase matched
signals for wave mixing experiments on polyatomic molecules, and we look forward to this and other applications in the future.

\section{Acknowledgments}

This collaboration was primarily funded by the Scientific Discovery through Advanced Computing 
(SciDAC) program of  
the Advanced Scientific Computing 
Research, Basic Energy Sciences, Biological and Environmental Research, High Energy Physics, Fusion 
Energy Sciences, and Nuclear Physics programs of the U.S. Department of Energy, Office of Science.
Work performed at Lawrence Berkeley National Laboratory was additionally supported by the US 
Department of Energy Office of Science, Basic Energy Sciences program, 
contract DE-AC02-05CH11231, and work at the University of California 
Davis was supported by US Department of Energy contract No. DESC0007182.
We thank the National Energy Research Scientific Computing Center (NERSC) for computational resources.


\appendix

\section{Derivation for different electron one and electron two bases\label{diffsect}}

It is wasteful to define orbitals all in the same basis extending over the entire molecule.  
Many electrons, notably core electrons, will be localized.  To account for this, 
it is imperative to define orbitals on different grids with different spatial extent and resolution.  
Such a treatment then calls for Slater determinants belonging to different
classes containing different numbers of electrons occupying 
orbitals belonging to different grids.  However, since the one electron 
bases are not combined there is no problem of linear dependence
nor any significant issue related to orthogonality.

One could, for example, interpolate the density on the sparser grid onto the finer grid, then
 use $T^{-1}$ for the finer grid to evaluate the integral.  However, it is interesting
to try to adapt the derivation directly to the case of two different bases.

The derivation with two different bases for electrons one ($ij$) and two ($kl$) follows.
We define 
\begin{equation}
\label{eq:yklx}
y^{kl}(\vec{r}_1) = \int d^3\vec{r}_2 \ \chi^{(2)}_k(\vec{r}_2) \chi^{(2)}_l(\vec{r}_2) \frac{1}{|\vec{r}_1 - \vec{r}_2|}
\end{equation}
Applying the Laplacian to both sides of equation Eq.(\ref{eq:yklx}) and 
approximating $y^{kl}(\vec{r}_1)$ as a linear combination of the basis functions in $\vec{r}_2$,
\begin{equation}
\label{yklexpandx}
y^{kl}(\vec{r}_1) \approx \sum_{\mathrm{ALL} \ n} y^{kl}_n \chi^{(2)}_n(\vec{r}_1) \quad ,
\end{equation}
applying the Laplacian with respect to $\vec{r}_1$ to this expansion,
multiplying through by $\chi^{(1)}_m(\vec{r}_1)$, integrating over $\vec{r}_1$ and equating the results leads to
\begin{equation}
\sum_n y^{kl}_n T_{mn} = 2\pi \int d^3\vec{r}  \ \chi_{k}^{(2)}(\vec{r}) \chi^{(2)}_l(\vec{r}) \chi^{(1)}_m(\vec{r})
\label{reseqx}
\end{equation}
where $T_{mn} = -\frac{1}{2}\langle \chi^{(1)}_m | \nabla^2 | \chi^{(2)}_n \rangle$ are the kinetic energy matrix elements. 

Again using the resolution of the identity to approximate the density (sum of squares of basis functions),
the right hand side of Eq.(\ref{reseq}) is evaluated as
\begin{equation}
\sum_n y^{kl}_n T_{mn} = 2\pi \Delta_{(2)}^{-3/2} \delta_{kl} S_{km} 
\end{equation}
with $S$ the overlap matrix
\begin{equation}
S_{lm} = \int d^3\vec{r} \ \chi^{(2)}_l(\vec{r})\chi^{(1)}_m(\vec{r})
\end{equation}
giving
\begin{equation}
y^{kl}_n = 2\pi \Delta_{(2)}^{-3} \delta_{kl} \sum_m S_{mn} (T^{-1})_{mk}.
\end{equation}
Using these coefficients and inserting Eqs.(\ref{yklexpandx}) into 
\ref{eq:twoelectronsimple} gives the expression for the two-electron matrix elements
\begin{equation}
\label{eq:twoelectronexpandx}
[ij | kl] = \sum_n y^{kl}_n \int d^3\vec{r}_1  \ \chi^{(1)}_i(\vec{r}_1) \chi^{(1)}_j(\vec{r}_1)  \chi^{(2)}_n(\vec{r}_1).
\end{equation}
Once again, the integral on the RHS is evaluated by a resolution of the identity, this time in the $r_1$ density,
to obtain the expression
\begin{equation}
[ij | kl] = 2\pi(\Delta_{(1)}\Delta_{(2)})^{-3/2} \delta_{ij}\delta_{kl} \sum_{mn} S_{in} S_{mk} T_{nm}^{-1}.
\end{equation}
Because the electron one and electron two grids are not commensurate, more than 
one column (equivalently, with different indexing, one row) of $T^{-1}$ will have to  be stored. 

\bibliography{sincdvr.bib}

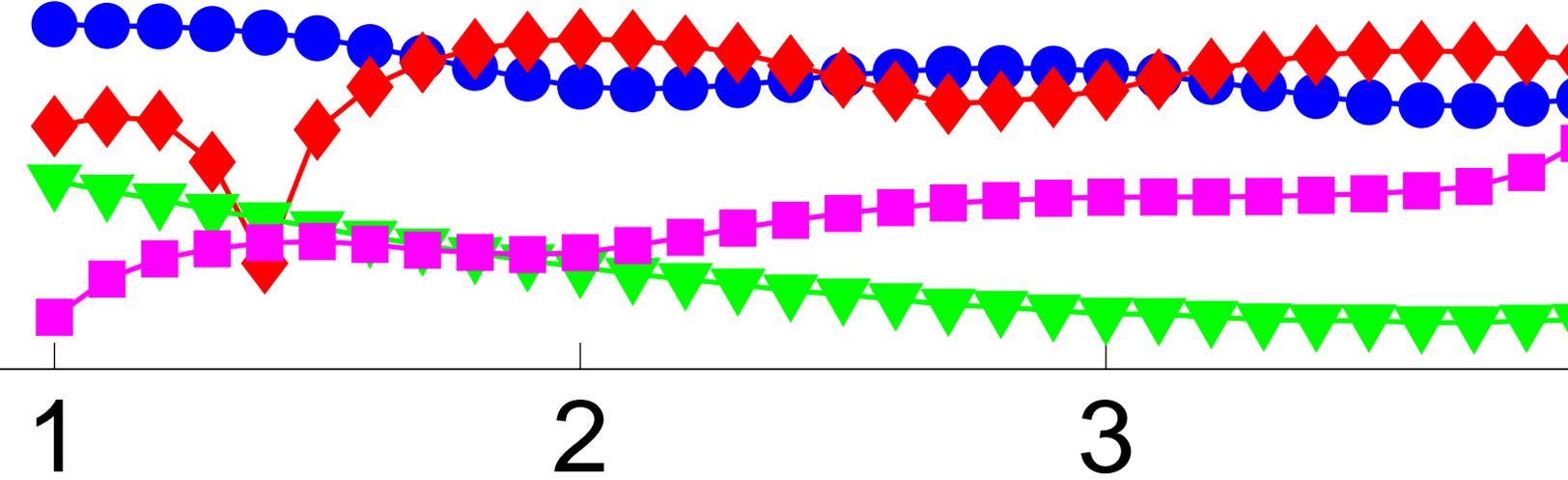
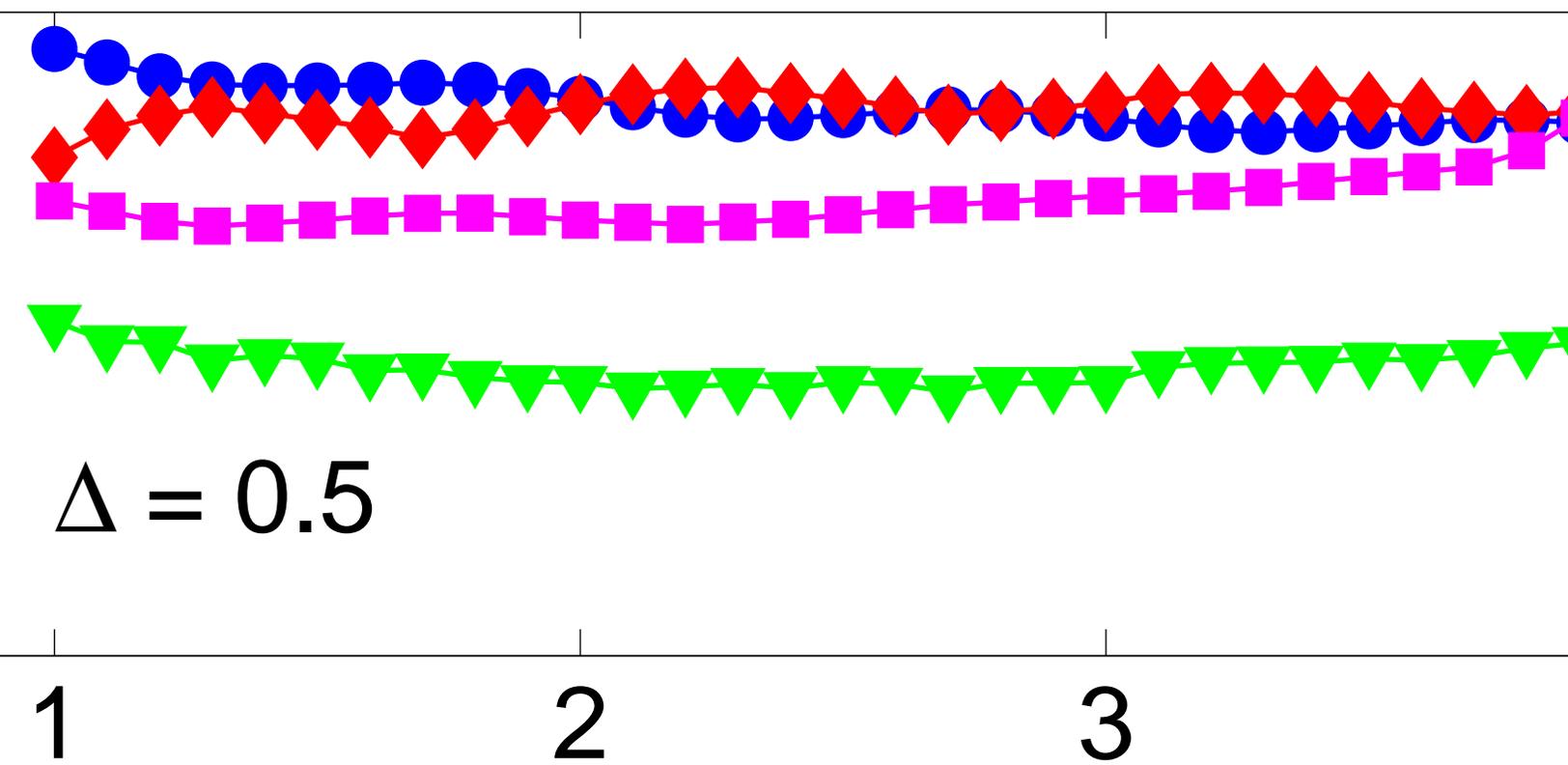
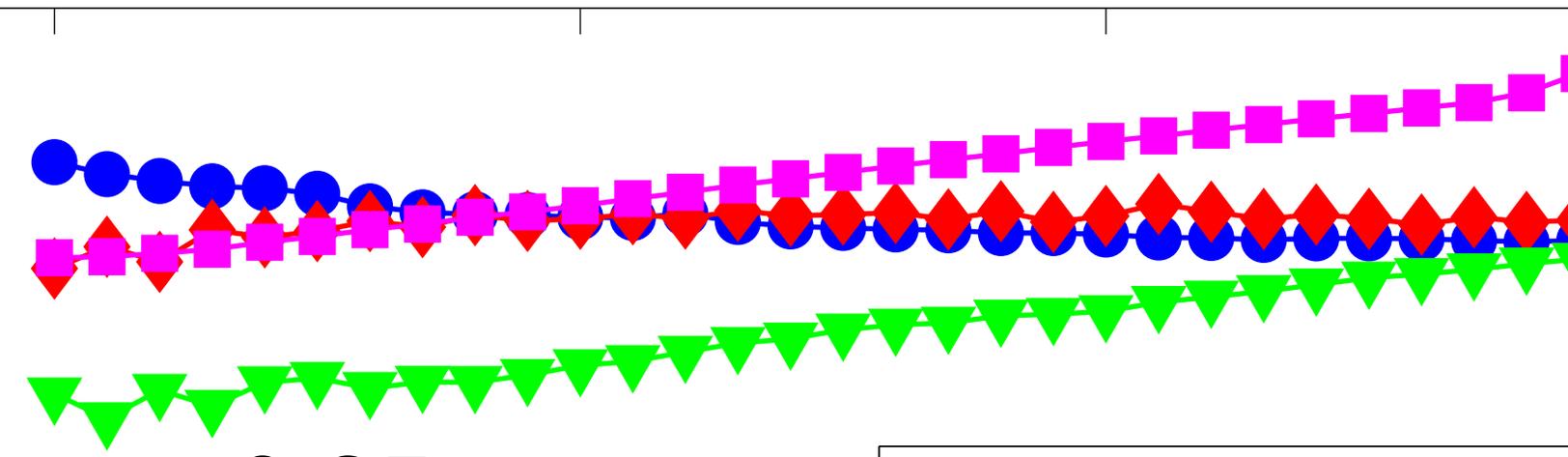
\end{document}